**Title**

A Conceptual Framework for Using Machine Learning to Support Child Welfare Decisions

**Author Names and Affiliations**

Ka Ho Brian Chor, Ph.D.[+][*][a] ([bchor@chapinhall.org](mailto:bchor@chapinhall.org))

Kit T. Rodolfa, M.P.P., Ph.D.[+][b] ([krodolfa@andrew.cmu.edu](mailto:krodolfa@andrew.cmu.edu))

Rayid Ghani, M.S.[b] ([ghani@andrew.cmu.edu](mailto:ghani@andrew.cmu.edu))

[+] Authors contributed equally to this manuscript

[*] Corresponding author

[a] Chapin Hall at the University of Chicago

 1313 E. 60th St.

 Chicago, IL 60637

[b] Heinz College of Information Systems and Public Policy

 Machine Learning Department

 Carnegie Mellon University

 5000 Forbes Ave.

 Pittsburgh, PA 15213

**Acknowledgments**

This work was partially funded by the University of Chicago-Chapin Hall Joint Research Fund.

**Declaration of Interest**

None

**ABSTRACT**

Human services systems make key decisions that impact individuals in the society. The U.S. child welfare system makes such decisions, from screening-in hotline reports of suspected abuse or neglect for child protective investigations, placing children in foster care, to returning children to permanent home settings. These complex and impactful decisions on children's lives rely on the judgment of child welfare decisionmakers. Child welfare agencies have been exploring ways to support these decisions with empirical, data-informed methods that include machine learning (ML). This paper describes a conceptual framework for ML to support child welfare decisions. The ML framework guides how child welfare agencies might conceptualize a target problem that ML can solve; vet available administrative data for building ML; formulate and develop ML specifications that mirror relevant populations and interventions the agencies are undertaking; deploy, evaluate, and monitor ML as child welfare context, policy, and practice change over time. Ethical considerations, stakeholder engagement, and avoidance of common pitfalls underpin the framework's impact and success. From abstract to concrete, we describe one application of this framework to support a child welfare decision. This ML framework, though child welfare-focused, is generalizable to solving other public policy problems.

# INTRODUCTION

## Child Welfare Decisions

At its core, the U.S. child welfare system can be conceptualized as a series of decisions the government makes in the best interest of children regarding their safety (e.g., Is the child's safety threatened at home?), well-being (e.g., Is the child thriving cognitively, behaviorally, emotionally, socially, and physically"), and reconnection to a permanent home setting outside the child welfare system (e.g., Can the child thrive and reunify safely with biological caregivers?). These decisions represent sequential and consequential milestones in children's involvement with the child welfare system. First, when a reporter such as a concerned neighbor or teacher calls the 24/7 abuse or neglect hotline, intake staff need to decide whether to initiate a formal child protective investigation or an alternative response, such as a community service referral. Second, if the decision is to initiate an investigation, a child protective service investigator identifies the types of alleged abuse or neglect and determines whether to substantiate individual allegations and the overall investigation. Third, if the investigator substantiates the investigation, the court system determines whether to remove the child from home and place the child in the legal custody of the responsible child welfare agency. Fourth, if the court removes the child from home, the child welfare agency determines where to place the child, for example, in a foster home or a residential facility. Finally, the child welfare agency monitors its temporary custody of the child to determine whether to return the child to a permanent setting, such as reunification with the child's family, adoption, or guardianship.

Significant constraints contextualize these child welfare decisions. Most child welfare agencies are underfunded, understaffed, and overburdened (Pecora, Whittaker, Barth, Borja, & Vesneski, 2018). Overwhelming caseloads permeate all decision milestones in terms of number of hotline calls received by hotline intake staff, number of investigations assigned to investigators, and number of foster care cases assigned to caseworkers. Limited time to make these decisions may also compromise the objectivity and quality of these decisions. Hotline intake staff might be legally bound to respond to each hotline call as soon as possible; investigators might only have limited time to conduct a thorough investigation; caseworkers might have to make an emergency placement change for a child whose existing placement is no longer viable. Decisionmakers also face the challenge of synthesizing relevant and available data to inform their decisionmaking. Data scarcity is critical at the hotline and investigation phases because the system knows little about the child at these junctures. Yet, the most important decisions—investigating a family and removing a child from their home—are made on this frontend. By the time a child enters the foster care system, prior decisions cannot be undone despite richer data available to improve our understanding of children's lives in foster care.

For decades, the serious nature of these child welfare decisions—both on tasking decisionmakers to make the "right" decision and to protect children's best interest—has called for actuarial tools or decision-support algorithms to ensure these decisions are expedient, data-driven, unbiased, and evidence-based (Drake, Jonson-Reid, Ocampo, Morrison, & Dvalishvili, 2020). Rather than replacing and automating the existing human-driven decisionmaking process, analytical tools that highlight risks and important details from historical data can help decisionmakers augment their expertise. A systematic, human-centered review of these algorithms used within the U.S. child welfare system (Saxena, Badillo-Urquiola, Wisniewski, & Guha, 2020) finds that implemented algorithms across decision milestones are disparate and wide-ranging from consensus-based assessments to predictive statistical methods. Rarely does a child welfare agency develop a coordinated, coherent framework for using decision support algorithms. Recent implementation of decision support algorithms illustrates the absence of such a framework. Often a child welfare agency emphasizes one decision in isolation because of local interests or in response to external mandates such as a consent decree, a lawsuit, or legislation. For instance, the Family First Prevention Services Act (FFPSA; P. L. 115-123) emphasizes identifying and serving children at "imminent risk" of entering foster care, thereby increasing child welfare agencies' attention to the frontend of the decision milestones. The Illinois Department of Children and Family Services contracted with an external firm to mine historical data to identify investigated children at risk for serious injury or death within two years, though this model was discontinued due to poor performance of the proprietary algorithm (Jackson & Marx, 2017). Partnering with university researchers, the Allegheny County Department of Human Services in Pennsylvania (Chouldechova, Putnam-Hornstein, Benavides-Prado, Fialko, & Vaithianathan, 2018; Cuccaro-Alamin, Foust, Vaithianathan, & Putnam-Hornstein, 2017) and the Wisconsin Department of Children and Families (2014) have developed and implemented predictive risk models to support screening decisions of reported child maltreatment and to identify children at risk of re-entering the child welfare system, respectively.

**Machine Learning as a Decision Support Tool in Child Welfare**

Machine learning (ML) is a family of statistical methods that "learn" from complex data over time and detect underlying patterns to support various tasks, including outcome predictions (Drake et al., 2020; Ghani & Schierholz, 2020). Although the specific methods might vary and range from regression to random forest to deep learning (Kim, 2018), most applications of ML leverage methods that allow more flexibility (and often complexity) in learning the underlying data patterns. Operationally, ML learns from experience (e.g., historical child welfare data and outcomes) with respect to some class of tasks (e.g., predicting child maltreatment) and performance measure (e.g., prediction accuracy), to improve that task performance with experience (e.g., improved prediction over time) (Mitchell, 1997). Predictive analytics is a related term that encompasses ML and other methods that predict an outcome a given model is trained on; predictive risk modeling often specifies the use of the prediction in terms of risk scores for each case (Drake et al., 2020). Child welfare agencies have begun to explore the extent to which ML tools might be able to improve their processes and outcomes for children involved (Russell, 2015). ML can benefit child welfare decisions when it is conceptualized, designed, evaluated, and implemented empirically and responsibly (Drake et al., 2020), taking into account the following considerations in the child welfare context.

## 1. Child Welfare Decisions are High-Stakes Decisions

Child welfare decisions in the aforementioned decision milestones have direct impact, punitive or assistive, on children and families involved. Punitive child welfare decisions concern potential deprivation of children's opportunities to grow up in their biological families and of parents' opportunities to self-correct their parenting behaviors. Assistive child welfare decisions concern protection of children's safety and opportunities to thrive outside a harmful environment. The use of prediction tools in these high-stakes decisions (Shroff, 2017), regardless of the methods involved, will influence decisionmakers to initiate an investigation or to extend a child's stay in foster care. The gravity of these decisions means that errors, risks, benefits, and consequences of using decision support methods, including ML, are especially costly, and therefore must be pre-specified, operationalized, and evaluated.

## 2. Need for Transparency in the Process, Data, and Methods Used

Decision support methods such as ML must be transparent and understandable to justify using prediction to support high-stakes child welfare decisions (Drake et al., 2020). A transparent model that clearly explains its predictions helps ensure that child welfare decisionmakers and the child welfare system are accountable to children, families, and the general public. A cautionary example of the contrary in which an ML model was a proprietary black box led to harmful results, such as over-classifying children as at high-risk of severe harm while missing those who actually experienced severe harm (Jackson & Marx, 2017). A transparent model facilitates the process for critique, evaluation, refinement, and improvement over time (Drake et al., 2020).

## 3. Child Welfare Context and Decisions are Ever Changing

Decision supports in child welfare often focus on using historical data to predict target events and behaviors in the future. These events (e.g., Will a child be maltreated?) and behaviors (e.g., Will a child successfully reunify with their biological family?) are, by definition, not static, objective phenomena. Although the problem of changing definitions due to changes and idiosyncrasies in policy and practice over time is not unique to the child welfare context, it means the detection and measurement of child welfare events and behaviors are imperfect. For example, a neighborhood over-surveilled by police may trigger a greater volume of hotline calls on abuse or neglect and subsequent investigations. But this does not necessarily mean the neighborhood has a higher prevalence of actual abuse or neglect than another neighborhood in which abuse or neglect is simply not reported. Similarly, definitions of abuse or neglect change over time and records could be expunged depending on individual state statutes. Using ML to support child welfare decisions must recognize these changing outcome context and definitions, via clear operationalization of the prediction of interest and capacity for managing the risks.

## 4. Variability in Availability and Quality of Information for Child Welfare Decisionmaking

Decision supports, especially ML, benefit from the availability of rich, extensive historical data. Child welfare systems generally do not lack administrative data because the legal and fiscal responsibilities of caring for children entail detailed tracking of all stages of child welfare involvement. Further, federal mandates require child welfare data collection in the State Automated Child Welfare Information System (SACWIS) and Comprehensive Child Welfare Information System (CCWIS). Unfortunately, the breadth and depth of child welfare data also incur burden on data collection and variability in data quality from one caseworker to another, and from one child welfare agency to another (Font, 2020). Structural and implementation variation means child welfare data may not be comparable or generalizable across time. Further, the quality of any prediction is only as good as the quality of the data input (Weyerer & Langer, 2019). An unintended consequence of decision supports built upon child welfare data is the exacerbation of institutional bias in the child welfare system (Drake et al., 2020). The level of racial, ethnic, and socioeconomic disproportionality and disparity consequent to child welfare decisions often becomes greater the further children are involved in system as compared to children in the general population (Child Welfare Information Gateway, 2016). Thus, ML decision supports using child welfare information must select predictors or features with care, and set up the training and validation of ML models with respect to time-dependent parameters.

## A CONCEPTUAL FRAMEWORK FOR INTEGRATING MACHINE LEARNING INTO CHILD WELFARE

The distinct decision junctures of child welfare involvement and the evolution of decision support algorithms are ripe for development of a conceptual framework that integrates ML approaches to support child welfare agencies in making informed and better decisions (Russell, 2015). We offer a comprehensive framework that allows researchers, practitioners, and decisionmakers to understand and determine where and how ML can improve key child welfare decisions as proxies for improving child outcomes, while building on existing predictive risk modeling frameworks (e.g., Drake et al., 2020). Our framework spans three phases that include seven steps (Figure 1): **Phase I: Problem identification and scoping:** (1) child welfare decision identification for improvement and decision scope; (2) data readiness; **Phase II: ML specification and development:** (3) model formulation; (4) model development; (5) model selection, evaluation, and understanding; **Phase III: ML deployment and maintenance:** (6) model deployment and field trial, and (7) model maintenance and monitoring.

All phases and steps have embedded ethical considerations as well as regular engagement with stakeholders including decisionmakers, caseworkers, and the families impacted. In practical terms, the framework guides how child welfare agencies might conceptualize a concrete decision that ML can help inform; vet available administrative data for building ML; formulate and develop ML specifications that mirror the target population and real-life actions the agencies are undertaking; deploy, evaluate, and monitor ML as child welfare context, policy, and practice change over time. Ethical considerations, stakeholder engagement, and avoidance of design pitfalls underpin the framework's impact and success (Drake et al., 2020).

Figure 2 briefly illustrates how Phases I, II, and III are connected, focusing on ML design decisions and definitions regarding the goal, outcome or label, cohort, decision frequency, evaluation metric, and action the model will inform. Each aspect of the workflow is described in more detail below. During Phase I, a high-level scope and problem statement is defined with the involvement of multiple stakeholders, including child welfare agency administrators, caseworkers, families, and ML practitioners. Here, we illustrate the steps focusing on reducing the number of children who disrupt from their foster home placement  (goal and outcome or label), such that every month (decision frequency), children currently in foster home placement (cohort) are identified for their risk of placement disruption (evaluation metric) to help caseworkers prioritize referrals to placement stabilization services (action). Phase II maps this high-level problem statement to specific design decisions required to address the problem with an ML model—the ML formulation bridges this divide between nebulous goals and well-specified parameters that can be expressed in data and code. This phase includes operationalization of the cohort (how foster home placement is defined), outcome or label (how placement disruption is defined), and model evaluation metrics in data terms (how risk of placement disruption is defined). Phase III describes how the ML system will be deployed, operationalized in caseworker workflows for placement stabilization services (action), and longer-term impact on foster care placement disruption in the child welfare system.

**Figure 1.** Conceptual framework for using machine learning (ML) to support child welfare decisions.

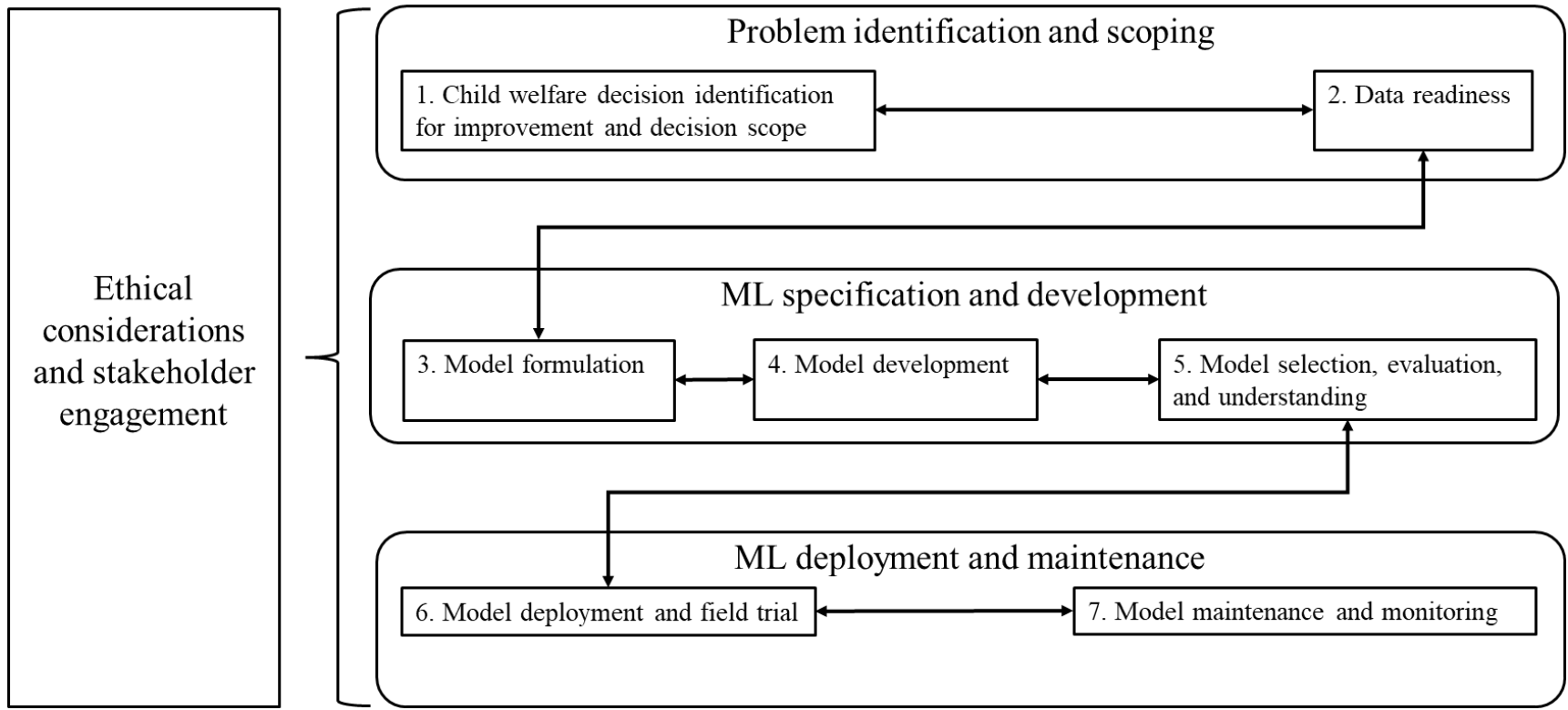

**Figure 2.** An example of how Phases I, II, and III are connected, focusing on ML design decisions and definitions regarding the goal, outcome or label, cohort, time, metric, and action.

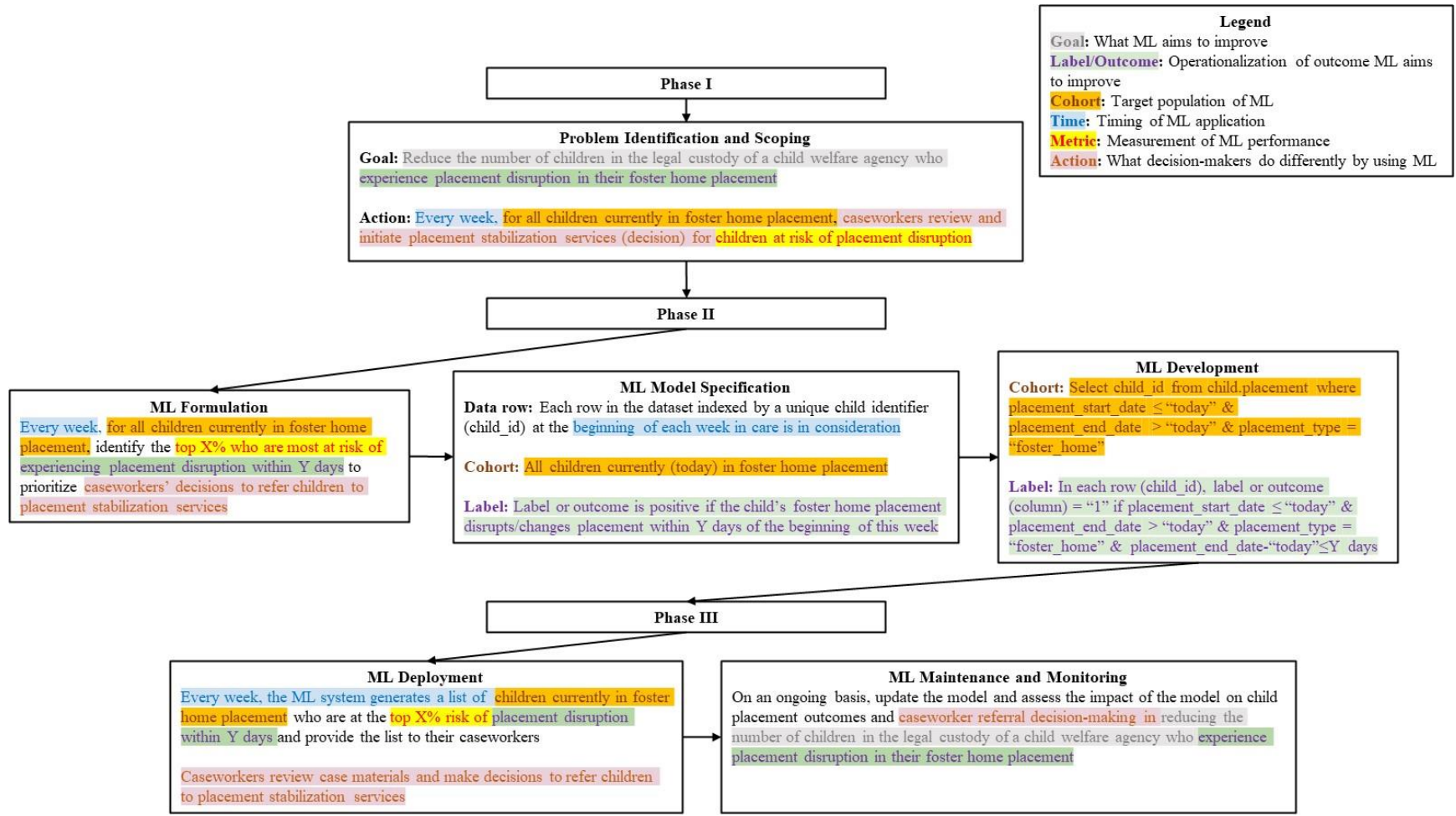

**Phase I: Problem Identification and Scoping**

A child welfare agency that contemplates using ML first needs to articulate the specific child welfare decisions it seeks to inform and improve. This initial phase requires the agency to operationalize the who, what, when, where, and how of the decisions, defining the high-level goals of the project, available actions or interventions to meet those goals, and how a data-driven system could inform those actions. This broad scope—established before getting into the details of modeling methods and data—ensures that the subsequent design decisions in Phases II and III are anchored in the social and the agency's goals.

*1. Child Welfare Decision Identification for Improvement and Decision Scope*

*a. What is the decision?* The target decision reflects a child welfare agency's priority for improvement and solving a specific problem. It might be to help hotline intake staff screen in calls to initiate investigations, or to help caseworkers identify discharge or aftercare resources.

*b. What outcomes will improve by improving this decision?* A child welfare agency should articulate a vision for the desired outcome from improving the decision with ML. For example, a desired outcome of screening in investigations with greater accuracy might be that hotline intake staff increase their efficiency and quality of screen-in for children who experience a level of threat that warrants investigations. A desired outcome of informing placement decisions for children in need of residential care with greater accuracy might be that children's length of stay in care is reduced, and that their placements become needs-based rather than capacity-based.

*c. When and how often is the decision being made?* A child welfare agency should identify or define the frequency of the decision and the event that initiates the decision process. For example, 24/7 hotline calls can come in any time of a day to which hotline intake staff from different shifts must respond; once assigned an investigation, child protective investigators might limited days to conclude an investigation. Other decisions can be anticipated by decisionmakers. Caseworkers who identify the need for a child's placement change might be expected to initiate such decisions within a known timeframe as governed by policies and procedures. The event that initiates the decision process must be clearly defined (e.g., upon receipt of hotline calls, at discharge planning meetings). It cannot be too broad (e.g., any time before a child enters or exits the child welfare system), vague (e.g., help hotline intake staff or caseworkers "do a better job"), or unmeasurable (e.g., using dates of observation notes or risk assessments filed in a binder).

*d. Who makes the decision?* A child welfare agency should define the scope and operational context within which the desired child welfare decision can be supported by ML. This includes clarifying the specific decisionmakers who can benefit from ML predictions. Key personnel might include frontline decisionmakers such as hotline intake staff, child protective investigators, and caseworkers, though more nuanced roles such as supervisors and administrators, depending on a specific child welfare agency, should also be considered.

*e. Who is impacted by the decision?* A child welfare agency should describe the population impacted by the child welfare decision. Generally, children are directly impacted by the decision—children in the general population being reported on abuse or neglect, children being investigated by child protective services, and children being placed in child welfare legal custody. Broadly, however, families and caregivers of the children are also impacted consequent to the decision made on behalf of the children by decisionmakers. The people impacted by the decision benefit from greater specificity in the identification process. For example, after a child is removed from home, a child welfare agency that is interested in early level-of-care planning on optimal placement within the first 30 days of removal might equip caseworkers with ML prediction on day 20. This arrangement translates to a target population of children who have been removed from home for 20 days, from whom data up to day 20 could be used to develop and implement a ML model to support caseworkers. The child welfare agency needs to know its resources and capacity to intervene on those impacted by the decision. For example, if hotline calls are screened out, are there primary prevention resources that might benefit children and families? If hotline calls are screened in, how quickly can investigators be assigned?

2. *Data Readiness*

*a. Are there sufficient and relevant data available at the time the decision is made?* At minimum, a child welfare agency must have relevant and reliable data on the outcome or dependent variable (i.e., a target child welfare decision or a target outcome consequent to the decision, such as child placement) and on a variety of related features or predictor variables (i.e., factors that contribute to the outcome or dependent variable). Although most child welfare agencies are data-rich, data availability varies greatly depending on the decisionmaking juncture at which the agency aims to support using ML. There are generally less data on children before they enter the child welfare system in the hotline or investigation phase than after their entry in the in-care phase. Children in the general population who have had no prior involvement with a child welfare agency have the least available data. Ideally, data are stored in machine-readable form in database; of high quality with limited incorrect or missing data; get integrated across different sources, providing an integrated view of the child; are refreshed such that the most recent data are available at the time of decisionmaking; and data ownership, security policy, privacy and confidentiality limitations, and accessibility guidelines exist. The Data Maturity Framework developed by the Center for Data Science and Public Policy (2016) provides a more detailed discussion of data readiness across these dimensions.

*Phase I Ethical Considerations and Stakeholder Engagement*

Ethical considerations in this phase include articulating the rationale for selecting a child welfare decision that needs ML support, be it for improving efficiency or fairness of the decisions. For example, the agency might realize the false positives from existing screen-in rate is above an acceptable threshold, which might unnecessarily traumatize children and families, and overburden hotline staff and investigators. Thus, the agency might aim to use ML to optimize screen-in rate. Stakeholder engagement in this phase seeks input from decisionmakers whose work is directly impacted by the integration of ML (e.g., hotline staff), to identify the primary problem area of existing practice (e.g., overwhelming caseloads) and refine the scope and operational context (e.g., assistance with identifying high-risk calls). It also involves engagement with indirect individuals at the child welfare agency, including supervisors, administrators, and leadership to ensure buy-in of ML at all levels, clarify the vision of using

ML, and scope out potential policy and procedural changes as a result of using ML. Most importantly, children and families directly impacted by a child welfare decision must have the opportunity to be empowered to express feedback on perceived burden on their lives and changes to their future. This effort can be supported by community and advocacy organizations that understand the basic rights, "ins and outs" of child welfare involvement, as well as the appropriate questions to ask, such as when and how a risk model score will be used for or against a child or a caregiver.. Other engagement might occur in child and family team meetings in which key decisions such as service, treatment, and placement planning are being made, and ML could play a role. Ethical considerations and stakeholder engagement should involve content and technical experts (e.g., quality assurance, information technology) regarding data security, confidentiality, strengths, and limitations of the agency's data systems that feed into ML.

**Phase II: ML Specification and Development**

In this phase, the problem identification and scope developed during Phase I are mapped onto the technical details of building an ML model, with a focus on ensuring that a well-formulated ML system will meet the specific needs of the child welfare agency. Much of this process involves a series of technical decisions that inform the structure and content of the data used as an input for developing, evaluating, and selecting an ML model. ML input data are structured as a matrix (Table 1)—that is, simply a flat file of typically numeric values specifying the unit of analysis (rows in the input data), features or predictors (covariates or independent variables), and label (target, outcome, or dependent variable). Design decisions on training and validating the ML system should also align with the corresponding structured data (Figure 3).

*3. Model Formulation*

    *a. What is the unit of analysis?* The unit of analysis in operational terms is what a row in

the data matrix represents (Table 1). It defines the cohort for analysis and consists of entities

(child, hotline call, family) at the time the decision is being made. For example, if a child welfare agency is predicting screen-in decisions for hotline calls, the cohort would be reported children within pre-specified temporal parameters (e.g., from January 1, 2015, through December 31,

2020). Thus, each "row" would consist of a unique hotline call for a specific reported child.

**Table 1.** An example of a mock data table for developing an ML system.

| child_id | decision_date | child_age | prior_placement_move | additional predictors | outcome |
|----------|---------------|-----------|----------------------|-----------------------|---------|
| 12345678 | 1/5/18 | 5 | 4 | ... | 0 |
| 74857346 | 2/20/18 | 5 | 8 | ... | 1 |
| 62345632 | 3/16/18 | 6 | 10 | ... | 0 |
| … | … | … | … | ... | … |

*b. What and when is the outcome or label of interest?* Child welfare outcomes as "ground

truths" (e.g., true child abuse) are approximated by proxy data. The typical outcomes in child

welfare use case scenarios are a binary classification of whether an event will occur in the future

(e.g., hotline staff will screen-in calls for investigations within an hour, children will enter care

within six month). Outcome operationalization must be sensitive to the changing definitions of

child welfare outcomes over time. The timing of the outcome must also be specified. Short-term

**Figure 3.** An example of defining training and validation sets of an ML system. outcomes might be more amenable to decisionmaking and interventions (e.g., screening-in high-risk abuse cases to minimize further abuse in the short-term) than long-term outcomes (e.g., screening-in high-risk abuse cases to connect families with preventive services to decrease future child removal from home).

*c. What should the ML system produce as an output?* The output of the ML system

**Figure 3.** An example of defining training and validation sets of an ML system.

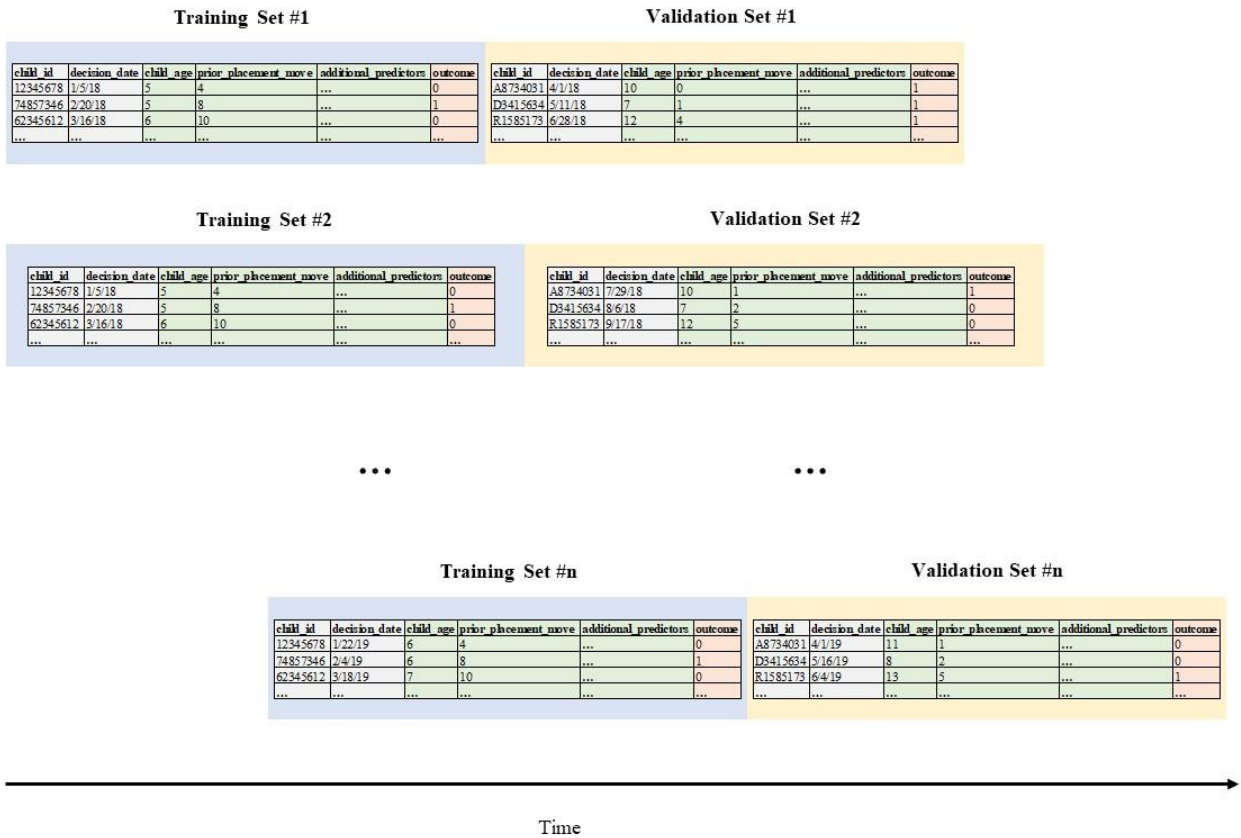

should match the existing output a child welfare agency uses to inform future actions. For example, if an existing process provides each caseworker with a list of homes they should visit next week, the ML output should provide a list of homes prioritized by a score or a rank-ordered list generated by an ML model. Another example might be that every month a child welfare agency generates a list of 100 children at highest risk of placement disruption for referral to placement stabilization services that can get distributed and allocated across caseworkers for review and interventions. Clarity of ML outputs is essential to developing a system well-suited to the agency's needs. The best performing model at selecting the 100 highest-risk children for an intervention might differ from a model best at selecting the 500 highest-risk children.

*4. Model Development*

*a. What are the features or predictors?* Features or predictors reflect local knowledge of a child welfare agency and are formed by raw data that become usable for modeling. For example, the agency might have a hotline call database that includes several potential data elements for predicting screen-in decisions. Each predictor must be further operationalized regarding aggregation (e.g., number of prior investigations), timespan (e.g., in the past year), imputation (e.g., mean, constant), and other criteria (e.g., prior abuse allegation vs. prior neglect allegation). The agency should initially scope out a broader set of data tables and data elements for further refinement (see data readiness in Phase I) as the ML development process unfolds. Because ML methods can handle a large number of features or predictors and collinearity, they benefit from casting a wider net rather than assuming the relevance of a narrower set of data elements. A common pitfall at this stage is unintentionally incorporating future information when making predictions on historical data. For example, a future child welfare entry date should not be used to predict substantiation of a prior investigation; clinical assessment fields that are updated and replaced rather than tracked longitudinally should not be used.

*b. How should training and validation sets be set up to develop and select ML models?* The goal of training and validating ML models is to mirror a child welfare agency's requirement for the selected models to perform well on new or unseen data when put into production to inform future decisions. Because these models typically seek to generalize to future decisionmaking, an appropriate strategy involves temporal validation (Ye et al., 2019), which provides a better evaluation of a model's expected performance than overall model fit criteria (e.g., $R^2$, Akaike Information Criteria, Bayesian Information Criteria) on the training set only. Thus, how training and validation should be set up reflects the temporal nature of a child welfare agency's use case. This replicates the cadence on which a model will be updated, timespan over which predictions will be made, and available historical information that can be used for model training at the time of prediction. The frequency of model updates should reflect the agency's capacity and resources for doing so, while the timing of prediction should reflect the use case of the decisionmaking process. Basic temporal parameters include specifying the earliest and latest date to include data on the child welfare decision or outcome of interest (e.g., hotline calls) and on the predictors (e.g., investigation history in the past year), and the timespan of the decision to predict (e.g., hotline calls in the next day or next week). Temporal parameters for model building include specifying the frequency of the prediction (e.g., monthly) in training and validating ML models. Specific "time-splitting" validation sets may be as simple and as conventional as the 70:30 rule for a single temporal cohort (i.e., use 70 percent of the earliest available data to train a model and the remaining 30 percent of the most recent data to test the model) or as complex as multiple temporal cohorts with varying training and validation periods over time (e.g., Figure 3).

*c. What evaluation metrics should be prioritized?* To ensure a model selected for deployment is well-suited to the problem at hand, target evaluation metrics and thresholds must be specified as related to the problem scope, constraints, and the priorities of a child welfare agency. Typically, the agency's motivation for using predictive analytics and predictive risk modeling is improved accuracy compared to existing practice (Drake et al., 2020). Suppose the agency has the capacity to serve 100 children at any given time in a specialized evidence-based program. The child welfare agency might focus on the metric of precision at the threshold of top 100 highest-risk children, which reflects the accuracy metric of interest (i.e., improved efficiency in the allocation of these limited resources) and the capacity of the child welfare agency to intervene on (i.e., top 100 highest model scores for defining a positive prediction of an outcome).

Fairness is another dimension of model evaluation, especially regarding existing processes that impact individuals with different protected class attributes differently. Due to concerns about existing disproportionality in the child welfare system (Child Welfare Information Gateway, 2016), race and ethnicity is often the most salient attribute in the evaluation of ML and predictive risk modeling context (Chouldechova et al., 2018). Fairness must be clearly defined. It might be in the form of setting an equal standard of predictive accuracy across subgroups of individuals, adjusting accuracy thresholds for different subgroups to overcompensate for known existing disparities between subgroups, or ensuring accuracy ratio between subgroups be proportional to the prevalence of the subgroups. Figure 4 is a "fairness tree" that guides the priority of fairness metrics (Rodolfa et al., 2020). This selection depends on the nature of the intervention as a result of the prediction (i.e., punitive vs. assistive as related to the target population), capacity of the intervention (e.g., most vs. some of the target population), specific subgroups of the target population (e.g., everyone regardless of actual outcome vs. regardless of actual need; everyone who is actually intervened vs. not intervened; everyone who does not need the intervention vs. everyone who needs the intervention), and consensus between stakeholders (e.g., child welfare agency vs. third-party purveyor).

Less measurable but also important is determining the agency's needs surrounding the interpretability of a deployed ML model. The agency might be motivated to "unpack" ML into its additive components that could explain the distinct contributions of protective or risk factors on which the agency can intervene. If so, seeing model coefficients or weights might be more desirable than aggregate risk scores. The definition of interpretability also depends on the end user capacity to translate and act upon ML results (Amarasinghe et al., 2020). Child welfare agency administrators and frontline caseworkers might need different ML-derived information to support their respective work. Whereas ML metrics that can be summarized at the child welfare system-level (e.g., proportion of cohort reaching a specific risk score threshold) might help administrators right-size resource allocation or service capacity, ML explanations that can transform important predictors (e.g., metrics for individual importance of features) into checklists or key questions might increase caseworker efficiency on day-to-day work.

*d. What types of ML models and corresponding specifications should be considered?*

Depending on the specific ML prediction problem at hand, there is a large array of available modeling methods (e.g., logistic regression, random forests, gradient boosting, etc.), each of which in turn has several hyperparameters over which to optimize (e.g., number of trees and their maximum depth in a random forest, regularization strength for a logistic regression, etc.). Because there is generally no prescribed a priori way to know what combination of modeling methods and hyperparameters will perform best for a given child welfare agency's problem, it is prudent to explore a large grid of methods and empirically evaluate their performance to choose the best option, unless there are resource or time constraints. This is true regardless of the child welfare decision of interest to predict. Unless there is a limitation or a strong preference for specific methods, the agency should consider a suite of methods from which to choose and compare findings. The relative merit and suitability of each method depends on the agency's priority. For example, the agency might value greater interpretability at the cost of sacrificing predictive accuracy, or vice versa, which could inform the selection of one method over another. Including a baseline method (e.g., simply ranking by one or two predictors or using heuristic decision rules) can serve as a benchmark for existing practice or what the agency could easily do in the absence of ML. This allows for a better evaluation of the relative improvements of using a ML model that can be weighed against the costs of deploying and maintain such a system.

*5. Model Selection, Evaluation, and Understanding*

This step takes all the models that have been specified and selects a subset of models to use in the future based on the trained ML models' ability to work well with unseen data, given a child welfare agency's general interest in informing future decisions. Evaluation metrics—predictive accuracy, fairness, interpretability—are compared across ML models and across time.

*a. How are ML models compared?* By casting a wide net with a sizable model grid, the agency will generate performance metrics from a large number of model specifications spanning several temporal training and validation sets, potentially yielding thousands of individual model results. To narrow down to a handful of model specifications for more in-depth exploration, the evaluation criteria (accuracy, fairness, interpretability) chosen in Phase II – Step 4C need to be compared across two dimensions: (1) aggregation of these metrics over time; and (2) the relative importance of each metric. For example, it may either be the case that recent performance is more relevant than more distantly historical performance (e.g., if policy or context has changed over time) or that performance across a long history is equally relevant (e.g., if context has been fairly stable). In the former case, a weighted average that puts more weight on recent performance may be appropriate. In the latter case, model selection might focus on averages of performance across the full set of temporal validation splits. Other aggregation strategies could focus on performance stability over time or how frequently a given model specification is "close" to the best possible performance across validation sets.

This process of model selection also needs to consider the degree of tradeoffs between the goals and needs of the agency regarding accuracy, fairness, and interpretability. Quantitatively weighing these different criteria against one another is generally difficult to implement. Instead, exploring these tradeoffs empirically based on the results from the grid of models run is often more tractable: How fair is the model specification that performs best on the chosen accuracy metric? How accurate is the model specification that performs best on the chosen fairness metric? What does the shape of the pareto frontier (among all the models that were run) across these metrics look like? Similar questions can be asked with regards to the level of interpretability of models (despite being less quantitative in nature). Likewise, this frontier can be expanded by including ML methods that seek to enhance model fairness or post-hoc model explainability. Ultimately stakeholders, presented with this range of measurable performance metrics, should inform the appropriate value judgment on which tradeoffs are more acceptable than others (e.g., sacrificing a nominal degree of accuracy for improving fairness) and justify the appropriate strategy to deploy given the needs and goals of the agency.

     *b. How are candidate ML models understood?* The model evaluation and selection process will often result in a handful of candidate ML models with similar performance across the various metrics of interest. Several analyses can help provide a better understanding of these models and their outputs, allowing for further narrowing this set of candidates to a specific. For example, how do these models compare with one another on the predicted score distributions, on the list of important features or predictors, or on the list of individuals with the highest predicted scores? If the candidate models are similar across these dimensions, the choice among them may be essentially arbitrary. If differences still exist, model deployment can become more complex, and may involve using recommendations from multiple models to generate a final list of individuals for intervention.

     Regardless of whether a single "best" candidate model emerges from the evaluation process, this phase of model understanding is critical for debugging and improving the model. The model should have a reasonable level of face validity: Do the important features or predictors make sense? Does the list of highest-score or highest-risk individuals make sense

based on historical data? Which features or predictors have the largest differences between higher-risk and lower-risk individuals? Is there any pattern of model error (e.g., false positive, false negative)? Even simple models (e.g., shallow decision trees) can perform reasonably well. Inspecting these simple models can help elucidate surprising features or predictors, as well as spurious correlations, issues with leakage, and other data patterns that may be present in more complex and better-performing models. Likewise, it is important to examine how these analyses change over time across temporal validation sets, as related changes in policy, practice, and the world, to gain an understanding of model robustness in a fluid environment.

*Phase II Ethical Considerations and Stakeholder Engagement*

Several areas of ethical considerations are pertinent in this phase, including transparency, privacy and security of the sensitive data being used, and defining, measuring, and improving the fairness of the models being built. With regards to transparency, every component of model formulation, development, training, validation, and interpretation should be documented and made accessible or public whenever possible. While the sensitive data about children and families cannot be made publicly available, providing well-documented, open-source code for the project along with other details about the goals and the scope of the effort may help promote public understanding and support. Accordingly, stakeholder engagement involves a review process with a child welfare agency and content and technical experts (e.g., child welfare, ML researchers) to refine specifications (e.g., time-splitting, predictors) as appropriate. Similarly, the choice of an appropriate conceptualization of fairness, let alone how it can be measured with statistical metrics, in a given policy context can be inherently difficult. This requires careful consideration of the potential harms associated with different types of errors and the perspectives of different stakeholders. Navigating this choice requires considerable discussion and input from these different stakeholders, as well as clear documentation and transparency around the decision itself. The how and why a particular fairness metric is selected should be questioned, explained, and justified from a child welfare agency's perspective, as well as from the stakeholders' perspective such that all involved understand the tradeoffs between predictive accuracy, fairness, and interpretability, and the concrete implications on decisionmaking such as casework burden and allocation of limited resources. Engagement with impacted children and families should focus on transparency of how ML models produce relevant outputs, such as a list of highest-risk children, and concrete implications of the outputs, such as more frequent caseworker contacts.

**Phase III: ML Deployment and Maintenance**

To understand how ML is applied in practice over time, the final phase involves the deployment and maintenance of a selected ML model.

*6. Model Field Trial and Deployment*

*a. How will a field trial be designed to validate that the ML system is worth deploying?* Because a well-designed ML model is based on historical data to make predictions within the historical context, the only way of truly assessing the applicability and generalizability of model performance is measuring it against future outcomes prospectively through a field trial. Further, such a field trial provides an opportunity to assess how well the model integrates into existing decisionmaking processes, such as how an ML model in practice is acted upon (e.g., how does caseworker behavior change as a result of knowing a child's risk score for placement disruption), overridden (e.g., how, when, and why caseworkers use their discretion instead of risk score recommendations), contributes to subsequent interventions (e.g., do caseworkers connect high-risk children with programs and services), as well as the effectiveness of the overall system from models through decisionmakers to interventions (e.g., do children in a child welfare agency experience less placement disruption as a result of ML implementation). A field trial could take the form of piloting the "best" ML model in a manageable context (e.g., in one administrative region), with a subset of caseworkers, or to support a subset of decisions in a limited timeframe, typically constrained by the need for statistical power to draw conclusions about the model's utility and impact. This means developing an implementation and maintenance plan, and a continuous quality improvement system that collects data on user feedback. The agency might also wish to conduct an assessment of organizational readiness regarding buy-in from decisionmaker staff, data collection staff, technical staff, and leadership.

**Figure 4.** Fairness tree, a guide for prioritizing fairness metrics (Rodolfa et al., 2020). FP/GS is the number of false positives divided by the total size of each group of interest (e.g., the subset of individuals of a given race, gender, etc), FN/GS is the analog with false negatives, FDR is the false discovery rate, FPR is the false positive rate, FOR is the false omission rate, and FNR is the false negative rate.

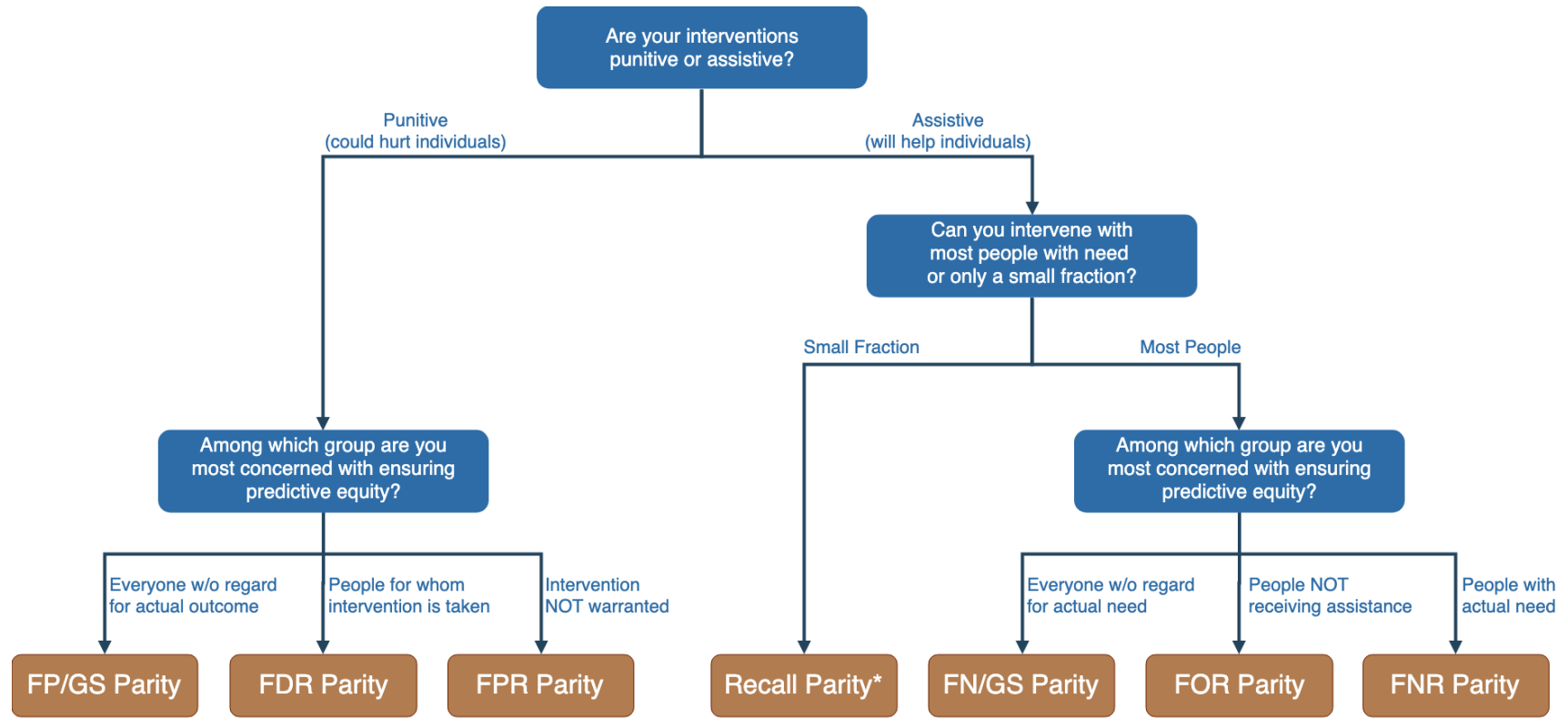

*b. Who is deploying the model?* A child welfare agency should clarify the actors of deploying the model. If the model is developed internally within the agency, the same personnel within the agency responsible for building the model can typically manage ongoing deployment. If the model is developed externally (e.g., with a university partner or contractor), the model needs to be handed-off to with clear specifications on how to deploy it. If a pilot is tested on a subset of decisionmakers (e.g., hotline intake staff in a "high-call" region), they should be trained on how the ML model works, what it means, and how they should incorporate the model into their decisionmaking practice.

*c. What and where are the technical resources for model deployment?* A child welfare agency will want to ensure ML deployment does not "break" and that if it does break, there are resources for troubleshooting. This would mean ensuring that the data and servers needed to run the model are ready, maintained, and provide sufficient computational resources for the model, allowing for any potential increase in scale with time. Often this conversation may also involve consideration of on-premise hardware vs. cloud-based infrastructure, and how these options fit into existing technical infrastructure and expertise. Additionally, the process for deploying the model, such as who is overseeing it and whether model predictions are automatically generated or manually updated, should be clearly defined.

*7. Model Maintenance and Monitoring*

Even after an ML model passes the deployment/field trial stage, it may degrade as the outcome definitions, context, policies, and world change. Metrics for model performance over time should be documented to observe consistency that might require systems intervention, for example, if new disparities, new program expectations, or new data arise. The model should be evaluated and monitored prospectively over time to identify performance anomalies relative to a baseline benchmark, and its impact on changing future decisionmaking behavior and subsequent children's outcomes.

*a. What are the metrics for monitoring?* Metrics for monitoring (e.g., predictive accuracy, fairness, interpretability) should correspond to metrics for evaluation in Phase II – Step 4C, which should align with a child welfare agency's priority, as well as provide measurements that capture when any of these important metrics change appreciably over time. Likewise, mechanisms for detecting sizable shifts in the data distribution, especially around outcome variables and important predictors, as well as in the predictors the model relies most upon can provide an early warning for a need to revisit model training and validation. The agency should develop tools, such as a dashboard or a user feedback system, for monitoring ML performance over time. Most importantly, the agency should track data on decisionmaker behaviors (e.g., caseworker referring high-risk children to an evidence-based intervention) as a result of deploying a model to ensure human application of the model does not perpetuate disparities (e.g., predominantly Black children are referred to the intervention). Similarly, the agency should track long-term outcomes of children impacted by these decisions (e.g., referred children benefiting from the program and exiting care). Collective trends on these metrics serve as triggers for retraining and revalidating models.

*b. Who is monitoring?* A child welfare agency should identify the responsible staff or unit for conducting monitoring, retraining, and updating ML models as necessary should degradation occur. This responsibility might fall under the purview of existing quality assurance or informational technology teams and processes or necessitate the development of new ones. This oversight should be supported by technical staff who can translate the technical aspects of ML systems to guide monitoring. Individual-level feedback on model performance might also require caseworkers to provide structured input on the relevance and use of the model in their work.

*c. Who is accountable to errors and resolving the errors?* Should decisionmaking errors with the use of ML incur tangible consequences to a child welfare agency (e.g., child death, delayed permanency), the agency should follow a protocol for recourse or corrective actions. This might mean further staff training on using ML information, fine-tuning ML models, and reporting these errors with specific follow-up resolutions that will be implemented and reviewed over time. In instances where traceable ML errors negatively impact the target population (e.g., a child missing out on receiving services), there should also be a procedure for recourse (e.g., additional outreach) in place to make mid-course corrections. Ideally this process is similar to the existing human decisionmaking process (if any) for reviewing and correcting errors.

*Phase III Ethical Considerations and Stakeholder Engagement*

Ethical challenges in this phase extend to how an ML model interacts with behavior, practice, and policy changes. A detailed implementation plan should accompany the deployment and field trial of a final ML model. This includes training staff to guide incorporation of ML outputs into their current work, update of a child welfare agency's policies and procedures to standardize the practice change, and continuous monitoring (e.g., following vs. overriding recommendations) and midcourse corrections of using ML (e.g., random auditing of cases or root cause analysis of errant data). While no decisionmaking system will be able to make perfect predictions about the future, regardless of whether it is entirely human-driven, entirely autonomous, or a hybrid of human and ML components, ensuring that mechanisms are in place to carefully consider prediction errors (e.g., when an opportunity for early intervention was missed), provides an opportunity to understand systematic patterns in these errors and allows for improving the decisionmaking system over time. A particularly salient risk with systems that involve ML models is the tendency to deploy them as "black boxes" that are not understood by agency staff and leadership, and are relied upon without proper monitoring to understand if model predictions continue to meet the requirements of accuracy and fairness in the face of a changing world. If a child welfare agency phases in ML implementation and begins with a pilot, site selection for the pilot should account for staff buy-in, capacity for change, and infrastructural supports. Stakeholders should be provided with a feedback loop to understand the impact of actual decisions made. The goal of stakeholder engagement is to solicit user feedback—caseworkers, technical staff, children and families—for continuous quality improvement and identify technical, process, and recourse challenges.

**EXAMPLE APPLICATION: SUPPORTING CASEWORKER'S DECISION TO REFER CHILDREN TO A PLACEMENT STABILIZATION PROGRAM**

While many child welfare decisions represent the chronicity of child welfare involvement (Figure 5), existing ML literature tends to focus more on children's entry to (Chouldechova et al., 2018) or exit from (Ahn, Gil, & Putnam-Hornstein, 2021) the child welfare system, but less on decision supports that could improve children's experience in care. Thus, it may be beneficial to provide an example of applying the framework developed above on how a child welfare agency might develop an ML system to support caseworkers' decisions to initiate placement stabilization for children in care. Appendix Table A1[1] describes in brief other applications of the framework.

**Figure 5.** Child welfare involvement, relevant child welfare decisions that can be supported by machine learning, and relevant studies using ML or predictive modeling to support child welfare decisions.

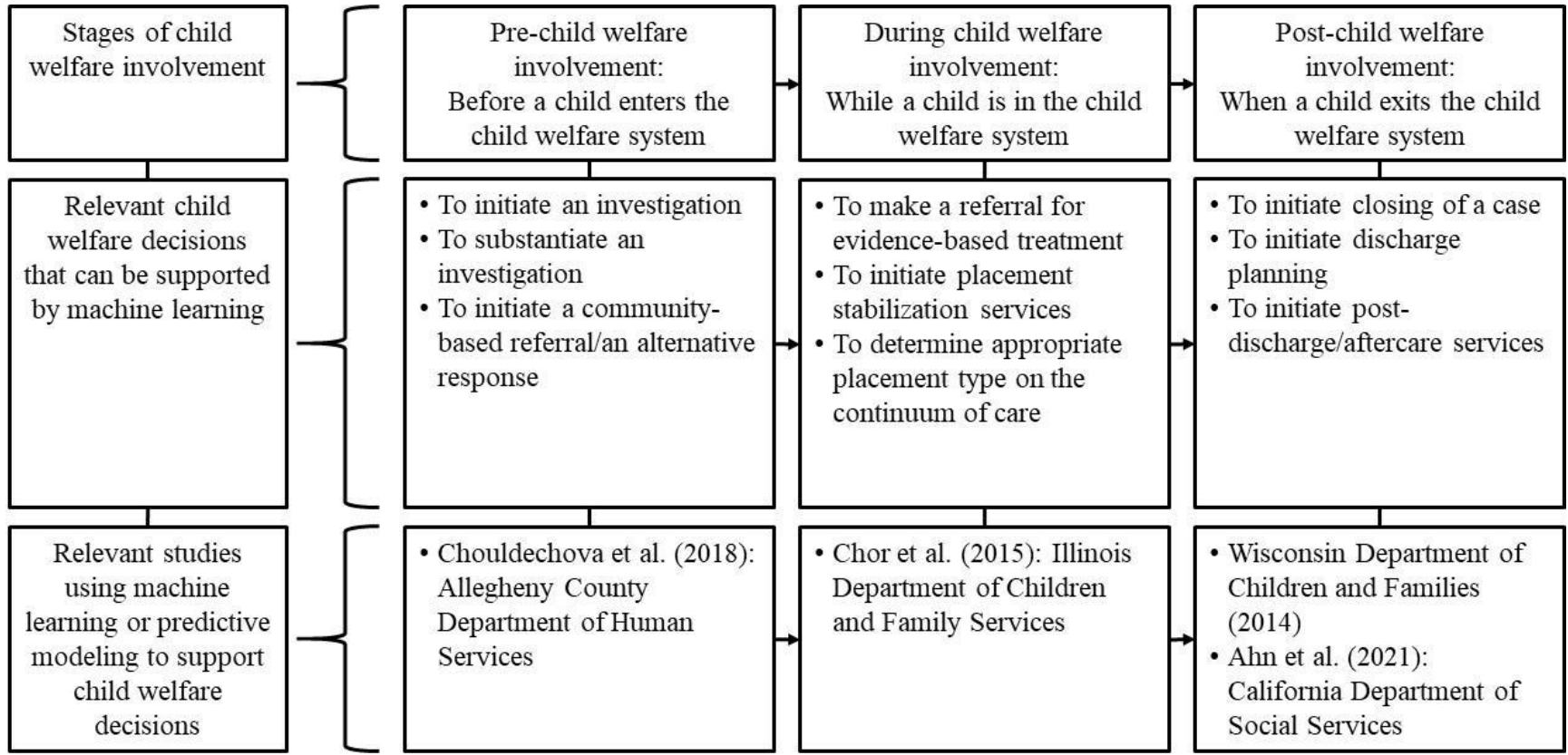

Placement disruptions in foster care can destabilize children's connections to the community, put children in vulnerable circumstances such as homelessness, and may result in children remaining in care longer than necessary or aging out of the foster care system (Dworsky, Wulczyn, & Huang, 2018). Seeking to improve children's placement stability, child welfare agencies provide various placement support interventions to children identified as being at risk of disruption. For example, in Illinois, the Department of Children and Family Services institutes a formal program called Clinical Intervention for Placement Preservation, which uses a multidisciplinary team model to preserve child and family social connections, provide, and linkchildren to targeted services for children at risk of placement disruption from foster or relative home settings (Illinois Department of Children and Family Services, 2013). However, limited intervention resources mean not all children can benefit from this program at all times. Further, limited staffing, competing priorities, and heavy caseloads mean caseworkers may not have the capacity for individualized case planning, which may leave them vulnerable to making ad-hoc, inconsistent, or even arbitrary decisions to refer children to this program. In this context, there is scope for applying placement decision support tools that previous work has suggested can improve well-being outcomes of children in care over time (Chor, McClelland, Weiner, Jordan, & Lyons, 2015). Because of these capacity and resource constraints, the agency is exploring ML to facilitate caseworkers' decisions to referral children with the highest needs to the program as timely and appropriately as possible. The agency envisions that an improved, anticipatory decision support process through ML can help caseworkers prioritize children in need of extra attention, stabilize and bolster current placement supports, and prevent placement disruption and institutionalization. Below we illustrate the application of the conceptual framework developed above to this specific use case.

**Phase I: Problem Identification and Scoping**

1. *Child Welfare Decision Identification for Improvement and Decision Scope*

   *a. What is the decision?* Whether or not caseworkers should initiate referrals of children on their caseloads to the placement stabilization program.

   *b. What outcomes will improve by improving this decision?* The primary outcome is improving the placement stability of children in care. ML supports for placement stabilization referral decisions could also improve the efficiency of referral decisions in the short-term, and afford caseworkers time for service planning, thereby allowing them to focus their limited time and resources where they could have the greatest impact on children.

   *c. When and how often is the decision being made?* Monthly, at the beginning of each month. Caseworkers can use decision supports to detect placement disruption flags to make referrals to the program.

   *d. Who makes the decision?* Caseworkers are responsible for making this decision about children on their caseloads. However, their decisions could be informed by consultations with other sources: the target children, supervisors, others involved in the care of the children, as well as existing decision support tools (e.g., checklists or other heuristics).

*e. Who is impacted by the decision?* Children in care are directly impacted by the decision. Collaterals, however, might include caregivers of children's current placements, children's treatment teams, and individuals with meaningful relationships with the children, all of whom could be affected by the consequences of initiating placement stabilization (e.g., separation from children, involvement with treatment). Caseworkers are also impacted as they might be required to document and justify these referral decisions in a timely manner.

*f. Who is impacted by the decision?* Children in care are directly impacted by the decision. Collaterals, however, might include caregivers of children's current placements, children's treatment teams, and individuals with meaningful relationships with the children, all of whom could be affected by the consequences of initiating placement stabilization (e.g., separation from children, involvement with treatment). Caseworkers are also impacted as they might be required to document and justify these referral decisions in a timely manner.

*2. Data Readiness*

*a. Are there sufficient and relevant data available at the time the decision is made?* Yes, the agency has historical administrative data for each child. These data elements are at the child-level to inform individual-level decisions and cover the entire child welfare history of the child. This allows ML to learn from historical patterns and be back-tested on historical outcomes (e.g., placement instability). Further, these administrative data pinpoint data elements that are relevant to the prediction use case at hand, including historical program referral decisions, placement disruptions, and clinical assessments of needs, meaning both the outcome of interest and potential predictors of this outcome are ascertained from the available data.

*Phase I Ethical Considerations and Stakeholder Engagement*

Stakeholder engagement at this stage includes gathering feedback from caseworkers about the existing challenges of placement stabilization as well as advocates representing the interests of children in care, and foster care providers to understand their perspectives on decisionmaking around placement stabilization. An important focus of these conversations is potential risks and harms (real and perceived) from these different perspectives. For instance, caseworkers might worry about whether such an ML system will undermine their autonomy and discretion in making decisions about how to support the children they are responsible for. Similarly overfocusing on metrics of placement stability per se might pose a risk of missing or delaying action in cases where a placement change is in the best interest of the child (e.g., a harmful foster home environment), highlighting the importance of understanding and communicating about how this decision support system fits into the broader goals of the agency, how it will be evaluated with respect to direct measures of child wellbeing, and how it fits into existing processes of measurement and evaluation (such as when adverse events do occur).

Privacy and data security are also important considerations at this stage, especially in a case where the ML system is being developed by an external partner with technical expertise in these areas. There, deidentified data can be shared and only accessed within a secure computing environment protected by appropriate access controls and usage monitoring, with these measures detailed in a data use agreement that clearly defines responsibilities of each partner and steps to be taken in the event of a breach.

**Phase II: ML Specification and Development**

*3. Model Formulation*

*a. What is the unit of analysis?* After consolidation of all relevant data elements pre-specified in Phase I, each "row" of the target data file will represent a unique child in care in a foster home setting at the beginning of each month to align with monthly decision supports. This directly implies that there will be multiple rows for each child, representing their current state at the beginning of each month they are in a foster home setting.

*b. What and when is the outcome label of interest?* Since placement stability is the ultimate outcome of interest, we operationalize that outcome by defining the outcome or label (a term often used in ML literature to define the outcome of interest or the target variable in modeling setup) to be "1" if a child experiences at least one placement move within a subsequent six-month period. Concretely, if a data row identifies a child in care as of January 1, 2020, then the outcome or label is "1" if they experienced at least one placement move between January 1, 2020, and June 30, 2020, and "0" otherwise.

*c. What should the ML system produce as an output?* Every first day of the month, the system should generate a list of 100 children with the highest risk of placement disruption for caseworkers to focus on. Note that the size of this list reflects the overall capacity of the program (i.e., 100 referrals). In addition to risk scores, caseworkers will likely also need more nuanced information such as characteristics of these high-risk children to customize service and placement planning, in addition to receiving supports in managing variations in the distribution of high vs. low-risk caseloads across caseworkers over time.

*4. Model Development*

*a. What are the features or predictors?* Features derived from the raw data sources described in Step #2 would focus on demographics as well as historical patterns of clinical treatment and interactions with child welfare services, such as prior placement changes, prior needs for placement stabilization services, clinical assessments across different dimensions, and past and ongoing clinical care. Each of these features would be aggregated over several historical windows relative to the date of prediction, for instance, capturing the number of placement changes in the past 1, 3, 6, 12, and 24 months, allowing the model to capture how changes and trends in historical patterns predict future outcomes. Although not immediately available, the agency may also explore obtaining data from the schools attended by children in their care (which would be structured into historical time-varying features on attendance, discipline, and academic performance) as well as court and law enforcement data to provide features on history of interactions with the criminal justice system.

   *b. How should training and validation sets be set up to develop and select ML models?*
Because the aim of the models is to generalize to support program referral decisions in the future, the training and validation splits should reflect the temporal nature of the problem. If the models are being re-trained every month (i.e., the same cadence on which predictions are being made), developing monthly training/validation set pairs would reflect the deployment setting. However, this approach could be computationally intensive while adjacent validation sets might provide little marginal information about model performance, as the children and their circumstances might change relatively little in monthly increments. As such, a reasonable starting point might be to sample training-validation sets every three months. In each pair, the training set would cover the time from the beginning of the data to six months before the date of the validation set (with the six-month gap reflecting the time needed to observe outcomes in the training set), the validation set would focus on all children in foster care on a given day and their outcomes over the subsequent six months. For example, if one training set includes children in foster care from the beginning of the available data through January 1, 2019 (with a subsequent six-month observation of placement stability outcome or label measured until June 1, 2019), then the associated validation set would include children in foster care as of June 1, 2019 (with a subsequent six-month observation of placement stability outcome or label measured until January 1, 2020). Note that a given child will be present more than once in this training set because the unit of analysis is not unique children but children who are in foster care at a point in time. Hence, the features associated with each of these child-level data rows would reflect the child's circumstances and case history as of the time of the date associated with the given row. How frequently dates are sampled within a given training set is also a balance between computational requirements and marginal performance gains, but every 4-8 weeks would likely be a good starting point in this context.

 *c. What evaluation metrics should be prioritized?* Given the limited capacity of the agency to provide placement support services to only 100 children each month, it is important that these limited resources are efficiently allocated to children who would benefit from them. In terms of ML accuracy metrics, this goal of efficient allocation maps to focusing on the precision (i.e., positive predictive value) in the top 100—that is, among the 100 highest-risk children the model recommends for the program each month, how many would in fact undergo a placement disruption in the next six months in the absence of the support services? Equally important to this efficiency goal is the agency's goals to ensure racial fairness of their allocation. Again considering the relative scarcity of their intervention resources, the agency might focus on ensuring children with need for additional supports from different racial and ethnic backgrounds have equitable access to the services, a notion of "equality of opportunity" that is captured by measuring disparities in recall (i.e., true positive rate) across groups (Hardt, Price, & Srebro, 2016; Rodolfa, Lamba, & Ghani, 2021; Rodolfa et al., 2020). Finally, because of the complex nature of child welfare needs and many factors which may contribute to a placement disruption (and, in turn, might be best served by different types of interventions), the agency would have some preference for more readily interpretable models—that is, if a model type for which explanations of its predictions could be easily provided (either inherently or using post-hoc methods) had very similar (or perhaps slightly lower) performance in terms of accuracy and fairness relative to a less interpretable method, the agency would be inclined towards deploying the former. Although no quantitative evaluation metric exists for "interpretability" per se, such models might both help caseworkers decide when to act on (or override) the model's recommendations as well as potentially providing useful information for intervention selection. For example, if the program offers services that target specific clinical factors that predispose children to placement disruptions, then it would be important that caseworkers see the breakdown of overall risk by contributing factors to assess for program fit.

*d. What types of ML models and corresponding specifications?* Although the technical staff at the agency might have more familiarity with regression methods such as logistic regression, they suspect more flexible, non-linear methods might be able to pick up on the complex relationships in child welfare data in predicting placement stability. As such, the agency chooses to cast a wide net over an extensive "grid" of binary classification methods (i.e., yes/no disruption outcome, yes/no referral) by considering a suite of model types (e.g., logistic regression, decision tree, random forest, gradient boosting, etc.) and hyperparameter options (e.g., regularization type/strength, tree and forest sizes, neural network architectures). In this way, the agency would understand potential model tradeoffs, consider options based on its prioritized level of performance, and consults with stakeholders.

*5. Model Selection, Evaluation, and Understanding*

*a. How are ML models compared?* After running the large grid of models described above, the ML developers will have results from hundreds of model specifications spanning several temporal training/validation sets, yielding thousands of individual model results. To narrow these down to a handful of model specifications to explore in more depth, they will need to combine the evaluation criteria (accuracy, fairness, interpretability) chosen in Phase II – Step 4C across two dimensions: (1) aggregation of these metrics over time; and (2) the relative importance of each metric. Here the child welfare agency considers a program where policies and practices have been relatively stable over time. The agency might choose to average performance metrics for each model specification across temporal validation sets. If, however, policies or the child welfare context had changed considerably over time, the agency might choose to instead put more weight on performance on more recent validation sets. Likewise, observed variability in model performance across validation sets might suggest a focus on aggregation methods that penalize model specifications with a high standard deviation across sets. Further, because it is difficult in this case to have an a priori agreement on how to quantitatively weigh goals of accuracy, fairness, and interpretability relative to one another, the ML developers instead choose to plot these different metrics against one another, allowing them to consider a pareto frontier of possible tradeoffs across models, from which a menu of several different options can be drawn. These options are then brought before stakeholders and decisionmakers, allowing them to weigh models with different possible metrics against one another in more concrete and actionable terms. For example, models with simpler specifications, that are more interpretable, and that perform relatively well might be more appealing to stakeholders than more complex models that are only marginally more accurate.

   *b. How are candidate ML models understood?* This step involves a deeper dive into a manageable set of candidate ML models that have passed the initial model comparisons, such as the menu of options reflecting different tradeoffs across different program goals discussed above, as well as some model types showing similar performance on these metrics, to understand if this translates into similar recommendations for intervention. This deeper dive might start with reviewing overall predictor/feature importance or, for simpler models, directly inspecting the model itself, to first ask if the model's behavior appears to "make sense" or seems suspicious in ways that suggest additional debugging of the features or models is required. Here, input from the frontline caseworkers and program staff is invaluable to better interpret these results through the lens of their expertise and intuition. It is possible that some models are driven by predictors that are relevant to the program's eligibility (e.g., children with specific clinical needs) and case planning (e.g., historical placement instability, placement type) than other predictors less noticeable to caseworkers' understanding of the program (e.g., child's demographics). Since the agency equally values predictive accuracy, fairness, and interpretability, the tradeoffs in these evaluation metrics across candidate ML models could be discussed in open forums facilitated by ML developers with participation from caseworkers and child welfare administrators. A closer examination of children identified as high-risk/score and their characteristics should serve as gut-checks of model relevance for caseworkers familiar with the program. Any of these criteria that further differentiate among candidate models will inform a child welfare agency's decision to deploy a specific model. If this decision is not as clear cut by objective criteria, the agency might also consider keeping a handful of models and combining their recommendations, either for an initial field trial or even longer-term deployment.

*Phase II Ethical Considerations and Stakeholder Engagement*

The primary ethical consideration in Phase II is how the agency conceptualizes and measures fairness in this use case, as well as how it determines the relative importance of predictive accuracy, fairness, and interpretability to inform model comparisons and model selection. Determining the appropriate conceptualization of fairness can be challenging in any setting, especially those involving high-stakes decisions that affect people's lives, ultimately requiring input from different stakeholders, especially those who will be directly affected by the decisions being made. In the discussion above, we focused on recall (also known as true positive rate) disparity, which is often relevant in allocation of supportive interventions in resource-constrained settings and reflects the idea that placement stabilization interventions will be beneficial to the children who receive them such that errors of exclusion (missing children with need) involve a greater harm than errors of inclusion (benefitting the mass). However, from the perspective of foster and relative caregivers, these interventions might involve a burden of extra scrutiny or even involve a stigma. From this perspective, if errors of inclusion accrued disproportionately disadvantages some groups of children over others, fairness metrics that focus on false positives may be of concern. This is a particular concern regarding Black and Brown families being disproportionately screened-in for further investigations and whether predictive decision support tools perpetuate this self-fulfilling prophecy (Glaberson, 2019).

Additionally, there may be tension between models' performance in terms of accuracy, fairness, and interpretability, which can be equally challenging to determine the relative weight across these criteria. Again, active stakeholder engagement and communications are critical to this determination, which includes technical experts, caseworkers, and placement stabilization program staff. Technical experts are expected to offer complete transparency of ML types, specifications, and comparisons of measurable metrics; caseworkers are expected to explore the differential implications of different models being deployed (e.g., would they conduct even more information gathering? Would they be concerned about under- or overrepresentation of minority children in the program?); placement stabilization program staff are expected to offer feedback on the face validity of the models based on their on-the-ground understanding of program eligibility and strengths. Other ethical considerations include articulating the human and program costs of prediction errors (e.g., opportunity costs of offering a program slot to one child at the expense of another child who needs it more) and how models might be underused or misused (e.g., caseworkers completely ignoring or deferring to model recommendations).

**Phase III: ML Deployment and Maintenance**

*6. Model Deployment and Field Trial*

    *a. How will a field trial be designed to validate that the ML system is worth deploying?*

A field trial in this context would focus on two goals: (1) confirming that the model's predictions of placement stability generalize into the future as expected based on the performance on historical validation sets; and (2) understanding the effectiveness of the model in improving caseworker decisions and children's outcomes relative to the status quo processes by which placement stabilization supports are currently staffed, resourced, and allocated. Because the child welfare agency is concerned about changing usage of the program system wide, piloting of a final ML model provides a case study opportunity to examine how the model is identifying children for the program, how caseworker practice is changing, what implementation challenges might arise, and how bias beyond the ML model might manifest during implementation. Key steps precede piloting of the model. The child welfare agency should clarify the business process for integrating the ML model into existing casework and referral to the program. For example, caseworkers would need to know what ML information they will receive (e.g., risk score, importance features or predictors) for their caseloads, how (e.g., from an e-mail, in children's case records), when (e.g., first day of every month), and expectations on casework changes (e.g., use ML information to convene team meetings, update service plans, make referrals). These field trial considerations are highly dependent on the local child welfare context and culture because training caseworkers to change their standard care is a challenge in itself and is not specific to integrating ML. Thus, involving caseworkers to design and provide feedback on the pilot is as important as collecting outcome data on the "effectiveness" of the pilot.

*b. Who is deploying the model?* The commitment of the child welfare agency to using ML to change placement stabilization practice will dictate the primary actors and responsibilities of ML deployment. The hand-off of the ML model from technical experts to in-house agency staff would detail the procedures for running the model at the expected regularity (e.g., every month). Caseworker "champions" who are familiar with the purpose and use of the model might serve as peer resources or "expert trainers" for other caseworkers; university partners who have developed the ML model might transition into an ongoing ML expert role; information technology and quality assurance departments in the agency might liaise with university partners to provide ongoing data supports and oversight.

*c. What and where are the technical resources for model deployment?* The child welfare agency needs to ensure ML deployment does not "break" and that when it does break, there are resources for troubleshooting. This would mean ensuring that the data and servers needed to run the model are ready and maintained, and that the process for deploying the model, such as who is overseeing it and whether model information is automatically generated, is clearly defined. Caseworkers might need regular ML how-to trainings or refreshers so that the agency's investment in ML does not go ignored as another "new thing" that will not stand the test of time. The backend data chain that allows ML models to be developed, validated, evaluated, and compared might also need to be adjusted depending on data system upgrades.

*7. Model Maintenance and Monitoring*

*a. What are the metrics for monitoring?* Relevant ML monitoring metrics for placement stabilization should focus on model performance changes over time and measurable impacts of ML deployment. Metrics for ML impact are directly tied to the child welfare agency's focus on predictive accuracy, fairness, and interpretability. These metrics might include program referral volume, risk score distributions, changes in racial composition of referred children, children's placement stability outcomes, and caseworker feedback on benefits and challenges in using ML information. Ongoing upkeep of ML models will rely on monitoring of ML evaluation metrics to identify trends, anomalies, and other signals for retraining and revalidating models.

*b. Who is monitoring?* For continuity's sake, the child welfare agency might enlist the same technical resources and protocol for ML deployment as for ML monitoring. System-level monitoring by the agency would monitor metrics that indicate model performance over time and that when degraded, would trigger ML model refinement. Monitoring also includes on-the-ground monitoring of how the placements stabilization program is used following ML implementation. For example, do caseworkers feel pressured to over- or under-refer children to the program for different reasons? Is the program experiencing a shorter or longer waitlist? What are referred children's experience of the program? These stakeholder perspectives of monitoring should inform the technical refinement of ML models (e.g., additional features or predictors to consider, different evaluation metrics) to respond the changing program context.

*c. Who is accountable to errors and resolving the errors?* Errors concerning referral to placement stabilization may not be as well-defined as consequential decisions such as removal of children from their homes. Still, errors should be defined relative to program eligibility criteria (e.g., low-risk children referred to placement stabilization by mistake) or post-referral outcome benchmarks (e.g., referred children experiencing greater placement instability than non-referred children). Because it is possible that there is no existing recourse or consequence to caseworkers not referring high-need children to the program, mechanisms for recourse for these errors should be developed and might include further caseworker training and reviews of ML definitions for predicted outcomes. More generally, the agency should develop a plan for how the model will fit into existing processes for continuous quality improvement and evaluation of decisionmaking. For instance, when adverse events do occur involving children in care (e.g., running away from care), what, if any, review process currently takes place? When a model is introduced into decisionmaking, how will existing processes need to change to accommodate, understand, and improve predictions in the future?

*Phase III Ethical Considerations and Stakeholder Engagement*

Even the best ML model predicting placement stability or recommending referrals to placement stabilization does not address how the ML is used. Thus, ethical considerations in this phase focus on the development and review of a clear ML pilot, deployment, and monitoring plan. Roles, timelines, expectations, and benchmarks must be clearly defined. The best plan will only come to life provided that there is a clear feedback loop among all parties involved. This feedback loop would allow caseworkers to express feedback about ML without repercussions, technical experts to engage with caseworkers and vice versa, and children impacted or missed by the program to express their opinions.

**MISTAKES AND MISCONCEPTIONS TO AVOID**

In our experience building and deploying ML systems for a wide range of high-stakes policy decisions, we have encountered several common pitfalls and misconceptions. In the hope that it will help others successfully apply the framework developed above to their own work, we close our discussion by highlighting several mistakes and misconceptions that can arise in applied ML settings in child welfare.

## Forgetting ML Systems are Inherently Contextual

Because the context of each child welfare agency is unique, there is no "generic" or "all-purpose" predictive model that can simply be embedded in the agency's decisionmaking process. As illustrated above, an effective ML system needs to be contextual and designed for a specific population (e.g., children currently in placement), to predict risk of a specific outcome (e.g., placement instability) at a specific time horizon (e.g., six months) to support specific interventions (e.g., referral to a placement stabilization program) with a specific capacity (e.g., 100 children per month). A generic model that predicts "adverse" child welfare outcomes is not only likely to be ineffective but also capable of causing more harm than good.

## Misalignment between Evaluation Metrics and a Child Welfare Agency's Goals

Many "textbook" metrics used to evaluate models such as AUC, FI, or accuracy generally fail to appropriately capture the needs of real-world problems. For example, if an agency's goal is to identify 100 children to prioritize for certain interventions, the appropriate metric to use is precision (or positive predictive value) on the 100 highest risk children identified by the model. A model with the highest overall AUC will often not be the same model as the one that performs best on this more task-specific metric. Developing and using an evaluation metric that mirrors the agency's goals is critical to having a positive impact and buy-in.

## Insufficient Relevant Predictors

Prediction-focused ML models benefit from a wide range of predictors that allow them to pick up on complex historical patterns and relationships in the data. However, researchers and practitioners who are more familiar with causal inference and methods may be inclined to focus on a narrow set of predictors that are assumed to be more appropriate to those tasks. Taking a more expansive approach to a model's predictors upfront is often a better starting point with ML methods, which are well-suited to using a very large number of predictors with complex correlation structures (including highly multicollinear predictors).

## Over-focus on Model Fairness Instead of Process and Outcome Fairness

Too often appraisal of ML stops at evaluating model performance, accuracy, and fairness at the level of a model's predictions. Equally if not more important is how ML would change the resulting decisiomaking process (e.g., caseworker referral) and outcomes of those impacted by the changed process (e.g., children at risk of placement disruption). In short, a fair- or bias-free model does not inherently ensure that the actions of the decisionmakers or the resulting outcomes or interventions will be fair for different subgroups. As such, it is critical that a holistic perspective be taken to the performance and fairness of a child welfare agency as a whole, accounting for how ML is implemented and the impact on actual outcomes for those affected.

## Assuming Exclusion of Race from ML Undoes Systemic Racism

A common misconception in many applications of ML to socially salient projects is that merely leaving out sensitive attributes (e.g., race) from the model's feature or predictor set would guarantee fairness along that attribute, which is far from reality (Dwork, Hardt, Pitassi, Reingold, & Zemel, 2012). In child welfare, using a "color-blind" model might inadvertently result in screening-in investigations at a higher rate in predominantly Black neighborhoods in which proxies for structural racism such as poverty is associated with over-surveillance of child abuse and neglect. Simply leaving out sensitive attributes can just as readily make a model less fair as more (or have no effect whatsoever), making this well-intentioned approach somewhat misguided (Lamba, Rodolfa, & Ghani, 2021). Fairness of these systems must be measured and considered explicitly, which necessitates collection and use of sensitive attributes in the interest of auditing for, understanding, and taking steps to mitigate any biases that do exist (Drake et al., 2020).

**Overlooking Data "Leakage"**

Leakage in ML is using data to develop and train a ML model that it otherwise would not have access to at the point of prediction, validation, and deploying the model (Ghani & Schierholz, 2020). This is a common pitfall of working with administrative data (e.g., SACWIS or CCWIS) in applied settings that can lead to over-optimistic estimates of model performance and poor choice of models to deploy. Avoiding leakage requires a careful understanding of how and when source databases are updated (especially when fields are updated with new information), careful feature engineering to only use historical information, diligently respecting time in model validation, and field trials that test models on events that are truly yet to happen as of the time of model training.

**Treating Model Development as a One-time Endeavor and Forgoing Ongoing Monitoring and Evaluation**

Because child welfare administration, law, policy, practice, and the world change over time, the historical patterns upon which an ML model relies to make its predictions are likely to change as well. Without ongoing monitoring and periodic retraining, a deployed model's performance—in terms of both accuracy and fairness—is likely to degrade with time. Likewise, because no decision process is perfect (whether ML-informed or entirely human-driven), the continuous evaluation of mistakes made by the system is essential to tuning and improving its decisions. Especially when ML models are built by external experts, there can be a risk of a "black box" system simply getting deployed and used in a static fashion. Doing so risks ignoring problems that may arise in the system prospectively or at the time of deployment, making a plan for longer-term maintenance and monitoring (using either internal or external technical resources) a critical aspect of any ML system deployment.

**CONCLUSIONS**

This paper proposes a conceptual framework for using ML to support child welfare decisions. The goal of the framework is to provide multi-stakeholders—ML developers, child welfare and social science researchers, and child welfare administrators and practitioners—with a unified understanding of what, when, and how ML can support key stages of child welfare decisionmaking in the interest of improving the equity and effectiveness in children's outcomes. Navigating the building, testing, and deployment of a ML decision support system can be a daunting undertaking. This is especially true for a child welfare agency that may have less familiarity with these methods. This framework attempts to reduce the barriers to doing so by highlighting the discrete steps in the process and important decisions to be made at each step. Nonetheless, several aspects of ML and integration of ML in child welfare are beyond the scope of this paper. The specific ML model methods and their relative strengths and weaknesses are better described elsewhere in the literature (Ghani & Schierholz, 2020). Implementation of a well-designed, integrated ML system in a child welfare agency is likely unique to the agency's context. Implementation science literature addresses topics such as training of the child welfare workforce or building an agency's internal capacity to adopt, implement, sustain, and evaluate ML as part of the agency's standard operation (Church & Fairchild, 2017; Russell, 2015). Likewise, the specific decision points, goals, and organizational constraints will vary greatly with agency and context. This paper, however, offers broad guidance that is applicable to practitioners across wide-ranging settings, articulates key questions to address, as well as mistakes, and misconceptions to avoid regardless of the specific implementation strategies. The lessons embodied in this work reflect our experiences applying ML methods across different high-stakes policy settings, including child welfare decisionmaking. This paper will help others benefit from our experiences, avoid the many pitfalls we have encountered along the way, and apply state-of-the-art tools and technologies to improve the lives of some of society's most vulnerable children and families.

**APPENDIX**

**Table A1.** Other applications of the conceptual framework for integrating machine learning (ML) to support child welfare decisions.

| Framework | Stage of child welfare involvement | |
|---|---|---|
| | **Before a child enters the child welfare system** | **When a child exits the child welfare system** |
| **Phase I: Problem identification and scoping** | | |
| 1. Child welfare decision identification for improvement and decision scope | | |
|   a. What is the decision? | • Initiate an investigation based on a hotline report of child abuse or neglect | • Close a child welfare case |
|   b. What outcomes will improve by improving this decision? | • Reduction of adverse events for children that result from not initiating an investigation when doing so is warranted | • Reduction of adverse events for children that result from closing a child welfare case "too early" when ongoing support is still necessary |
| | • Reduction of stigma and intrusion to families that result from initiating an | • Reduction of adverse events that result from closing a child welfare |

| Framework | Stage of child welfare involvement | |
|---|---|---|
| | **Before a child enters the child welfare system** | **When a child exits the child welfare system** |
| | investigation of inaccurate allegations when doing so is unwarranted | case "too late" when doing so is unwarranted |
| | • Increase in efficiency of allocating limited hotline and investigation resources to warranted investigations | |
| c. When and how often is the decision being made? | • As hotline reports come in | • Every month at regular case review intervals per child welfare policies and procedures for discharge planning |
| d. Who makes the decision? | • Hotline intake staff | • Caseworkers |
| e. Who is impacted by the decision? | • Potential child victims of hotline reports | • Children in care who are expected to exit care |
| | • Families of hotline reports | • Families and caregivers to whom children will return |

| Framework | Stage of child welfare involvement | |
| --- | --- | --- |
| | **Before a child enters the child welfare system** | **When a child exits the child welfare system** |
| 2. Data readiness | | |
| *a.* Are there sufficient and relevant data available at the time the decision is made? | • Hotline intake data<br>• Prior child welfare records (if victims had prior child welfare involvement) | • Child welfare records associated with current in-care spell<br>• Prior child welfare records (if children had prior child welfare involvement) |
| Phase I Ethical considerations and stakeholder engagement | • Gathering decisionmaker (hotline intake staff, caseworkers, children) feedback on existing decisionmaking process for improvement<br>• Understanding goals and risks from different stakeholder perspectives<br>• Discussing potential existing biases in the child welfare system (e.g., over-surveillance in child protection, judicial processes for approving case closure) and options for mitigating them in the modeling process | |

| Framework | Stage of child welfare involvement | |
|---|---|---|
| | **Before a child enters the child welfare system** | **When a child exits the child welfare system** |
| | • Ensuring mechanisms are in place to protect privacy and security of sensitive data involved in the project, especially if outside collaborators are involved | |
| **Phase II: ML specification and development** | | |
| 3. Model formulation | | |
| a. What is the unit of analysis? | • Each "row" of data is a hotline report as of the time of the report | • Each "row" of data is a child in care as of the first day of a given month |
| b. What and when is the outcome or label of interest? | • Indicator of a screened-in hotline report (report date) that results in an investigation that becomes substantiated within three months | • Indicator of a child exiting care (case closing date) re-entering care (case opening date) within a year |
| | • Indicator of a screened-out hotline report (report date) that results in another report within three months | |

| Framework | Stage of child welfare involvement | |
|---|---|---|
| | Before a child enters the child welfare system | When a child exits the child welfare system |
| c. What should the ML system produce as an output? | • For each hotline call received, a risk score and an associated recommendation for whether to initiate an investigation (based on a threshold reflecting both the agency's capacity for investigation and the relative costs associated with errors of inclusion or exclusion). The score may either be provided in real-time or on whatever cadence investigation decisions are made (e.g., hourly or daily). | • Monthly risk score of future child welfare involvement and associated recommendation for whether to close the case for all children currently undergoing discharge planning |
| 4. Model development | | |

| Framework | Stage of child welfare involvement | |
| --- | --- | --- |
| | **Before a child enters the child welfare system** | **When a child exits the child welfare system** |
| a. What are the features or predictors? | • Prior hotline reports, investigations, and child welfare records relevant to risk assessment and safety planning , as well as structured and unstructured data collected by the hotline intake staff during the call itself | • Child welfare records relevant to aftercare planning (demographic, clinical, and placement characteristics) from throughout the child's current (and any previous) spell with the agency<br><br>• Potentially (if available) administrative data from interactions with other entities, such as school or criminal justice data |
| b. How should training and validation sets be set up to develop and select ML models? | • Weekly training-validation splits | • Monthly training-validation splits |

| Framework | Stage of child welfare involvement | |
|---|---|---|
| | **Before a child enters the child welfare system** | **When a child exits the child welfare system** |
| c.  What evaluation metrics should be prioritized? | • Precision (at the model's recommendation threshold) of screened-in reports resulting in substantiated investigations within three months<br>• False negative rate disparities by race and administrative region | • Precision at the model's recommendation threshold of closed cases re-entering care within a year<br>• Recall disparities by race and administrative region |
| d.  What types of ML models and corresponding specifications should be considered? | • Explore a wide range of ML models (logistic regression, decision trees, random forests, gradient boosting) to prioritize predictive accuracy<br>• For less readily interpretable models, explore usefulness of post-hoc explanations to help inform the model selection processs | |
| 5. Model selection, evaluation, and understanding | | |

| Framework | Stage of child welfare involvement | |
|---|---|---|
| | **Before a child enters the child welfare system** | **When a child exits the child welfare system** |
| a. How are ML models compared? | • High level of average precision at the score threshold across validation sets, with some weight on stability of performance over time<br><br>• Investigate potential tradeoffs across models with highest accuracy and fairness metrics for input from stakeholders | • If discharge criteria, policies, or resources are subject to change over time, may prioritize recent precision at the model's score threshold<br><br>• Additional weight on model interpretability (either inherently or through available post-hoc methods) as actionable may help prioritize resources for case closing and aftercare planning<br><br>• Investigate potential tradeoffs across models with highest accuracy and |

| Framework | Stage of child welfare involvement | |
| --- | --- | --- |
| | **Before a child enters the child welfare system** | **When a child exits the child welfare system** |
| | | fairness metrics for input from stakeholders |
| b. How are candidate ML models understood? | In addition to standard analyses (score distributions, feature importance, feature crosstabs), seek input from hotline intake staff and investigators to conduct case reviews of highest-risk children to ensure model-predicted aligns with expertise-predicted risk | • In addition to standard analyses (score distributions, feature importance, feature crosstabs), seek input from caseworkers and discharge programs to ensure highest-risk children's needs can be met by existing resources |
| Phase II Ethical considerations and stakeholder engagement | • Understanding the appropriate way to conceptualize fairness in each context and mapping to a statistical measure <br> • Determination of acceptable tradeoffs between predictive accuracy and model fairness | |

| Framework | Stage of child welfare involvement | |
| --- | --- | --- |
| | Before a child enters the child welfare system | When a child exits the child welfare system |
| | • Engagement with hotline intake staff and caseworkers to ensure they understand how models are developed and change their practice<br><br>• Articulation of costs of prediction errors and consequences of misusing models | |
| **Phase III: ML deployment and maintenance** | | |
| 6. Model field trial and development | | |
| a. How will a field trial be designed to validate that the ML system is worth deploying? | • Targeted pilot rollout with buy-in from hotline intake staff and caseworkers in regions/offices that are generally responsive to innovations<br><br>• Clear field trial plan with details on how existing decisionmaking (i.e., screening in hotline reports, closing child welfare cases) will incorporate ML and change | |
| b. Who is deploying the model? | • ML technical experts involved with developing and piloting of ML assume ongoing implementation role<br><br>• Hotline intake staff and caseworkers with experience in ML supports serve as peer resources | |

| Framework | Stage of child welfare involvement | |
|---|---|---|
| | **Before a child enters the child welfare system** | **When a child exits the child welfare system** |
| c.  What and where are the technical resources for model development? | • Child welfare information technology and quality assurance collaborate with ML technical experts on data sharing and linkage<br>• Depending on the resources and practices of the agency, deployment may either take place using on-premise or cloud-based hardware, with appropriate security measures in place for data protection<br>• If a case management system is used (for either hotline reports or discharge planning), model outputs may need to be integrated with this system<br>• Periodic systemwide trainings and technical assistance for hotline intake staff and caseworkers<br>• Close communications between ML technical experts and child welfare agency on data system upgrades | |
| 7. Model maintenance and monitoring | | |

| Framework | Stage of child welfare involvement | |
|---|---|---|
| | **Before a child enters the child welfare system** | **When a child exits the child welfare system** |
| a. What are the metrics for monitoring? | • Precision of screened-in reports resulting in substantiated investigations within three months<br><br>• Recall disparities by race and administrative region<br><br>• Alignment between actual hotline intake staff's screen-in decisions and model recommendations<br><br>• Drift over time in model performance metrics, disparities, and important features | • Precision of closed cases re-entering care within a year<br><br>• Recall disparities by race and administrative region<br><br>• Alignment between caseworkers' case-closing decisions and model recommendations<br><br>• Drift over time in model performance metrics, disparities, and important features |
| b. Who is monitoring? | • ML technical experts on ML performance over time | • ML technical experts on ML performance over time |

| Framework | Stage of child welfare involvement | |
|---|---|---|
| | **Before a child enters the child welfare system** | **When a child exits the child welfare system** |
| | • Hotline intake staff to provide "on-the-ground" monitoring of screen-in practice changes | • Caseworkers to provide "on-the-ground" monitoring of changes in case-closing practice and aftercare planning |
| c. Who is accountable to errors and resolving the errors? | • Integration with existing processes for hotline report reviews and updates to policies or practices when avoidable adverse events occur (e.g., subsequent abuse or neglect to a child after a call was screened-out for investigation) | • Integration with existing processes for case reviews and updates to policies or practices when avoidable adverse events occur (e.g., subsequent abuse or neglect to a child after a case has been closed) |
| | • Additional trainings and feedback provided to hotline intake staff who are more prone to screening-in hotline | • Additional trainings and feedback provided to caseworkers who are |

| Framework | Stage of child welfare involvement | |
| --- | --- | --- |
| | Before a child enters the child welfare system | When a child exits the child welfare system |
| | reports that do not result in substantiated investigations | more prone to closing cases that result in re-entry |
| Phase III Ethical considerations and stakeholder engagement | • Clearly articulated pilot and deployment plan that is communicated to all responsible parties | |
| | • Field trial developed to understand relationship between fairness in model recommendations, actual decisions, and outcomes from interventions | |
| | • Plans and mechanisms for ongoing model monitoring and improvement, with particular focus on avoidable errors and disparities | |
| | • Develop and communicate mechanisms for feedback and recourse for those affected (both children and families) by the model-informed decisions | |